\documentclass{article} 
\usepackage{arxiv,times}

\usepackage{amsmath,amsfonts,bm}

\def\eqref#1{equation~\ref{#1}}

\def\1{\bm{1}}

\DeclareMathAlphabet{\mathsfit}{\encodingdefault}{\sfdefault}{m}{sl}
\SetMathAlphabet{\mathsfit}{bold}{\encodingdefault}{\sfdefault}{bx}{n}

\usepackage{hyperref}
\usepackage{url}

\hypersetup{
    colorlinks = true,
    allcolors = {purple},
    linkbordercolor = {white},
}

\usepackage{amsmath}
\usepackage{mathtools}
\usepackage{amssymb}
\usepackage{amsthm}
\usepackage[linesnumbered,ruled,vlined,norelsize]{algorithm2e}
\usepackage{bbm} 
\usepackage{tabularx}
\usepackage{longtable}
\usepackage{booktabs} 
\usepackage[dvipsnames]{xcolor}
\usepackage{enumitem}
\usepackage{subcaption}

\DeclareSymbolFont{bbold}{U}{bbold}{m}{n}
\DeclareSymbolFontAlphabet{\mathbbold}{bbold}

\usepackage{minitoc}

\usepackage{listings}
\usepackage{caption}
\usepackage[dvipsnames]{xcolor}
\usepackage[T1]{fontenc}
\usepackage{zi4} 

\definecolor{warmbase}{RGB}{253, 252, 248}   
\definecolor{warmkeyword}{RGB}{180, 60, 0}   
\definecolor{warmstring}{RGB}{160, 82, 45}   
\definecolor{warmcomment}{RGB}{100, 100, 80} 
\definecolor{warmnumber}{RGB}{120, 110, 100} 

\lstdefinestyle{mystyle}{
    language=Python,
    backgroundcolor=\color{warmbase},    
    commentstyle=\color{warmcomment}\itshape,
    keywordstyle=\color{warmkeyword}\bfseries,
    stringstyle=\color{warmstring},
    basicstyle=\ttfamily\footnotesize,
    numberstyle=\tiny\color{warmnumber},
    breakatwhitespace=false,         
    breaklines=true,                 
    captionpos=b,                    
    keepspaces=true,                 
    numbers=left,                    
    numbersep=5pt,                  
    showspaces=false,                
    showstringspaces=false,
    showtabs=false,                  
    tabsize=4,
    frame=single,
    framerule=0.5pt,
    rulecolor=\color{warmnumber!40},     
}

\newenvironment{longlisting}{\captionsetup{type=listing}}{}

\title{\textcolor{purple}{Policy-Conditioned Policies for Multi-Agent Task Solving}}

\usepackage{authblk}

\author[1]{Yue Lin}
\author[2]{Shuhui Zhu}
\author[3]{Wenhao Li}
\author[1]{Ang Li}
\author[1]{Dan Qiao}
\author[2,4]{\\Pascal Poupart}
\author[1]{Hongyuan Zha}
\author[1,4]{Baoxiang Wang}

\affil[1]{The Chinese University of Hong Kong, Shenzhen}
\affil[2]{University of Waterloo}
\affil[3]{Tongji University}
\affil[4]{Vector Institute}

\affil[ ]{ \texttt{\{yuelin,bxiangwang\}@link.cuhk.edu.cn} }

\iclrfinalcopy 

\begin{document}

\doparttoc 
\faketableofcontents 
\maketitle

\begin{abstract}
In multi-agent tasks, the central challenge lies in the dynamic adaptation of strategies.
However, directly conditioning on opponents' strategies is intractable in the prevalent deep reinforcement learning paradigm due to a fundamental ``representational bottleneck'': neural policies are opaque, high-dimensional parameter vectors that are incomprehensible to other agents. In this work, we propose a paradigm shift that bridges this gap by representing policies as human-interpretable source code and utilizing Large Language Models (LLMs) as approximate interpreters. This programmatic representation allows us to operationalize the game-theoretic concept of \textit{Program Equilibrium}. We reformulate the learning problem by utilizing LLMs to perform optimization directly in the space of programmatic policies.
The LLM functions as a point-wise best-response operator that iteratively synthesizes and refines the ego agent's policy code to respond to the opponent's strategy. We formalize this process as \textit{Programmatic Iterated Best Response (PIBR)}, an algorithm where the policy code is optimized by textual gradients, using structured feedback derived from game utility and runtime unit tests. We demonstrate that this approach effectively solves several standard coordination matrix games and a cooperative Level-Based Foraging environment.
\end{abstract}

\section{Introduction}

The study of multi-agent tasks, typically formalized as Markov games (or stochastic games)~\citep{shapley1953stochastic}, presents a complexity that fundamentally transcends single-agent decision-making. In these settings, the environment is non-stationary from the perspective of any individual agent, as the state transitions and expected rewards are co-determined by the evolving policies of interacting agents. Consequently, there is no single optimal policy in isolation; rather, an agent must continuously compute a \textit{best response} to the varying strategies of its opponents.

Canonical approaches to this problem, particularly in Multi-Agent Reinforcement Learning (MARL), typically rely on opponent modeling~\citep{he2016opponent}. Agents attempt to infer the current policies of their counterparts from the history of interactions. However, inferring complex, co-adapting policies from limited, high-dimensional, and non-stationary historical data remains an exceptionally difficult problem.

A theoretically more direct alternative is to explicitly condition an agent's policy on the \textit{actual} policy of the opponent. In the domain of game theory, this ideal is captured by the concept of \textit{Program Equilibrium}~\citep{tennenholtz2004program}, where agents delegate decision-making to programs that can read and condition on each other's source code. This framework theoretically enables rich cooperative outcomes (e.g., via ``mutual cooperation'' proofs in the Prisoner's Dilemma) that are inaccessible to standard Nash equilibria.

However, Program Equilibrium has remained largely theoretical and disconnected from modern learning approaches. In the prevalent Deep Reinforcement Learning (DRL) paradigm, implementing such policy-conditioning is practically intractable due to a ``representational bottleneck''. Policies are represented as deep neural networks defined by millions of opaque parameters. Inputting one neural network's weights into another is computationally prohibitive and semantically ill-defined, as the parameter vector is a high-dimensional, unstructured, and non-unique representation of behavior~\citep{goodfellow2016deep}.

The central thesis of this work is that we can overcome this bottleneck by leveraging Large Language Models (LLMs). We propose representing policies as \textit{programmatic policies}, which are executable, human-interpretable source code. Unlike opaque tensor parameters, source code is semantically dense and structured. By utilizing LLMs as approximate interpreters, we can operationalize the game-theoretic concept of Program Equilibrium within a learning-based framework. This allows agents to condition on the opponent's source code, enabling rich cooperative outcomes that are theoretically inaccessible to standard Nash equilibria but have previously been difficult to realize in practice.

This paradigm shift enables us to operationalize policy-conditional adaptation in a fundamentally new way. Instead of searching for a static policy via gradient descent on parameters, we shift the optimization landscape to the code space, where the LLM functions as a point-wise best-response operator. Given the source code of an opponent's policy, the LLM generates and iteratively refines the source code of the best-response policy.

We formalize this learning process as \textit{Programmatic Iterated Best Response (PIBR)}. This algorithm proceeds as an iterative dynamic, where at each step, an agent's policy code is optimized against the fixed code of its opponents. The optimization is driven by modern textual gradient techniques~\citep{yuksekgonul2024textgrad} using structured feedback, incorporating both (1) utility feedback from game trajectories and (2) unit test feedback to ensure syntactic and runtime correctness. By doing so, we present a practical instantiation of Program Equilibrium for general Markov games, demonstrating efficacy in several standard coordination matrix games and a cooperative grid-world task named \texttt{Level-Based Foraging}.


\subsection{Relation to Concurrent Work}
\label{Subsec: Concurrent Work}

During the preparation of this manuscript, a contemporaneous work titled ``Evaluating LLMs in Open-Source Games'' \citep{sistla2025evaluating} was presented at NeurIPS 2025. 
The work introduced a framework for Open-Source Games, evaluating how LLMs perform in game-theoretic settings when agents have full visibility of each other's source code.
We acknowledge that both the work and our work share a fundamental contribution: identifying that LLMs can serve as approximate interpreters to bridge the gap between theoretical Program Equilibrium and practical implementation. Both approaches utilize the transparency of code to enable conditional cooperation that is difficult to achieve with black-box neural policies. Chronologically, \citet{sistla2025evaluating} are the first to make this contribution.

However, our work diverges significantly from \citet{sistla2025evaluating} in motivation, methodology, and experimental scope.

\begin{enumerate}
    \item \textbf{Motivation from MARL Challenges:} While \citet{sistla2025evaluating} focuses on evaluating LLMs within a game-theoretic context, our approach stems fundamentally from the perspective of MARL. We identify programmatic policies specifically as a solution to the ``representational bottleneck'' and non-stationarity in MARL. We frame this not merely as an evaluation of LLMs, but as a necessary paradigm shift to solve the recursive reasoning challenges inherent in learning co-adapting policies.
    
    \item \textbf{Optimization via Textual Gradients:} \citet{sistla2025evaluating} primarily relies on the inherent reasoning capabilities of LLMs via prompting. In contrast, we propose \textit{Programmatic Iterated Best Response (PIBR)} as a formal optimization algorithm. Our method explicitly treats the LLM as a definable operator optimized via textual gradients \citep{yuksekgonul2024textgrad}. We incorporate a structured feedback loop that combines game utility with runtime unit tests, ensuring that the generated policies are not only strategically sound but also syntactically robust and executable.
    
    \item \textbf{Complexity of Tasks:} We extend the validation of programmatic policies beyond standard matrix games (which are the primary focus of concurrent works) to the \textit{Level-Based Foraging} environment. This is a complex grid-world task requiring spatial coordination and sequential decision-making, demonstrating the capability of our method to handle higher-dimensional state spaces and more intricate coordination dynamics.
\end{enumerate}

Acknowledging the precedence of \citet{sistla2025evaluating}, we refrain from submitting this specific version for formal conference publication to adhere to novelty standards. Instead, we release this manuscript as a technical report to establish the timeline of our independent contribution. We plan to incorporate these findings into future submissions that feature significant technical improvements or novel extensions beyond the scope of the current discussion.


\section{Preliminaries}

This section provides the necessary background for our method. We begin by establishing the formalisms for multi-agent learning, including game settings and solution concepts. Then we discuss the two fundamental challenges that motivate our work: non-stationarity and the complexities of opponent modeling. 
Finally, we introduce the concept of program equilibrium and provability logic, which provide the theoretical basis for agents that can condition their policies on the source code of their opponents.
A detailed discussion of related work is deferred to Appendix~\ref{Section: Related Works}.

\subsection{The Game}

In this work, we focus on the multi-agent task which is formulated as a \textit{Markov Game}~\citep{MarkovGame1994}, also known as a stochastic game~\citep{shapley1953stochastic}. 
A Markov Game for $N$ agents is defined by a tuple $\mathcal{G} = \langle N, \mathbb{S}, \{\mathbb{A}^i\}_{i=1}^N, \mathcal{T}, \{r^i\}_{i=1}^N, \gamma \rangle$.
$\mathbb{S}$ is the set of global states. $\mathbb{A}^i$ is the action space for agent $i$, and the joint action space is $\boldsymbol{\mathbb{A}} = \times_{i=1}^N \mathbb{A}^i$. The function $\mathcal{T}: \mathbb{S} \times \boldsymbol{\mathbb{A}} \to \Delta(\mathbb{S})$ is the state transition probability function. Each agent $i$ has a reward function $r^i: \mathbb{S} \times \boldsymbol{\mathbb{A}} \to \mathbb{R}$, and $\gamma \in [0, 1)$ is the discount factor.
At each timestep $t$, agents simultaneously select actions $a^i_t \in \mathbb{A}^i$. The system transitions to $s_{t+1} \sim \mathcal{T}(\cdot \mid s_t, \boldsymbol{a}_t)$ and each agent $i$ receives a reward $r^i_t = r^i(s_t, \boldsymbol{a}_t)$, where $\boldsymbol{a}_t$ is the joint action.

\subsection{The Solution Concepts}

The objective of each agent $i$ is to find an optimal policy $\pi^{i*}$ that, when combined with the policies of all other agents, maximizes its \textit{expected discounted return}. This cumulative reward, denoted $J^i$, is defined for a joint policy $\boldsymbol{\pi} = (\pi^1, \dots, \pi^N)$ as:
\begin{equation*}
J^i(\boldsymbol{\pi}) = \mathbb{E}_{\langle \mathcal{G}, \boldsymbol{\pi} \rangle} \left[ \sum_{t=0}^{\infty} \gamma^t r^i(s_t, \boldsymbol{a}_t) \right],
\end{equation*}
where $\mathbb{E}_{\langle \mathcal{G}, \boldsymbol{\pi} \rangle}[\cdot]$ denotes the expectation over the state transitions $\mathcal{T}$ and the agents' joint actions $\boldsymbol{\pi}$.

The primary solution concept is the \textit{Nash Equilibrium (NE)}~\citep{nash1950equilibrium}, a joint policy $\boldsymbol{\pi}^* = (\pi^{i*}, \boldsymbol{\pi}^{-i*})$ from which no single agent $i$ has an incentive to unilaterally deviate. Formally, $\boldsymbol{\pi}^*$ is a Nash Equilibrium if for all $i \in N$ and for all alternative policies $\pi^i$ available to agent $i$ ($\pi^i \in \Pi^i$), the following condition holds:
$$
J^i(\boldsymbol{\pi}^*) \ge J^i(\pi^i, \boldsymbol{\pi}^{-i*}).
$$
This general definition is refined based on the game's temporal horizon and the class of policies considered.

In the finite-horizon (episodic) setting, equilibria must hold at all decision points. The history $h_t$ is defined as $h_t = (s_0, \boldsymbol{a}_0, \dots, s_{t-1}, \boldsymbol{a}_{t-1}, s_t)$, and $\mathbb{H}$ is the space of all such histories.
One refinement is the \textit{Subgame Perfect Equilibrium (SPE)}~\citep{bielefeld1988reexamination}, which applies when players use history-dependent policies, $\pi^i: \mathbb{H} \to \Delta(\mathbb{A}^i)$, denoted $\Pi^i_{\mathbb{H}}$. An SPE $\boldsymbol{\pi}^*$ requires the NE condition to hold in every subgame, i.e., beginning from any possible history $h_t$. For all $i \in N$, $h_t \in \mathbb{H}$, and $\pi^i \in \Pi^i_{\mathbb{H}}$:
$$
J^i(\boldsymbol{\pi}^* \mid h_t) \ge J^i((\pi^i, \boldsymbol{\pi}^{-i*}) \mid h_t),
$$
where $J^i(\boldsymbol{\pi} \mid h_t) = \mathbb{E}_{\langle \mathcal{G}, \boldsymbol{\pi} \rangle} \left[ \sum_{k=t}^{\infty} \gamma^{k-t} r^i(s_k, \boldsymbol{a}_k) \mid h_t \right]$ is the expected future discounted reward given history $h_t$.
A second refinement is the \textit{Markov Perfect Equilibrium (MPE)}~\citep{fudenberg1991game}, which restricts policies to be Markovian and time-dependent, $\pi^i: \mathbb{S} \times \mathbb{T} \to \Delta(\mathbb{A}^i)$, denoted $\Pi^i_{\mathbb{M}}$. An MPE $\boldsymbol{\pi}^*$ requires the NE condition to hold at every state-time pair $(s, t)$. For all $i \in N$, $s \in \mathbb{S}$, $t \in \mathbb{T}$, and $\pi^i \in \Pi^i_{\mathbb{M}}$:
$$
J^i(\boldsymbol{\pi}^* \mid s, t) \ge J^i((\pi^i, \boldsymbol{\pi}^{-i*}) \mid s, t),
$$
where $J^i(\boldsymbol{\pi} \mid s, t) = \mathbb{E}_{\langle \mathcal{G}, \boldsymbol{\pi} \rangle} \left[ \sum_{k=t}^{\infty} \gamma^{k-t} r^i(s_k, \boldsymbol{a}_k) \mid s_t = s \right]$ is the expected future discounted reward from state $s$ at time $t$. Both SPE and finite-horizon MPE can be computed via backward induction~\citep{Farina2024_L16_StochasticGames}.

In the infinite-horizon discounted setting, a foundational result establishes that a Nash Equilibrium exists in \textit{stationary, Markovian policies}, $\pi^i: \mathbb{S} \to \Delta(\mathbb{A}^i)$~\citep{fink1964equilibrium, takahashi1964equilibrium}. Formally, a joint stationary policy $\boldsymbol{\pi}^* = (\pi^{1*}, \dots, \pi^{N*})$ is an equilibrium if for all $i \in N$ and for all alternative policies $\pi^{i\prime} \in \Pi^i_{\text{all}}$ (where $\Pi^i_{\text{all}}$ includes arbitrary history-dependent policies):
$$
J^i(\boldsymbol{\pi}^*) \ge J^i(\pi^{i\prime}, \boldsymbol{\pi}^{-i*}).
$$
This demonstrates that the stationary equilibrium policy $\pi^{i*}$ is robust against deviations to any non-Markovian strategy.

\subsection{Learning Methods for Markov Games}

The advancement of DRL has enabled the use of deep neural networks to approximate complex policies and value functions, greatly extending the applicability of RL to high-dimensional state and action spaces. MARL extends this DRL framework to environments with multiple interacting agents ($N > 1$). These agents may be cooperative, competitive, or operate in a mixed setting.

In a MARL setting, each agent $i$ learns its own \textit{policy} $\pi^i$. As defined in the previous section, this policy can be a mapping from the current state (a stationary, Markovian policy $\pi^i: \mathbb{S} \to \Delta(\mathbb{A}^i)$) or from the full history (a history-dependent policy $\pi^i: \mathbb{H} \to \Delta(\mathbb{A}^i)$).
In practice, policies are frequently history-dependent. Agents require history for two principal reasons: 
\textbf{(1)} First, in partially observable environments, an agent must utilize its history to infer the current underlying state. We assume a fully observable environment, so this point does not apply.
\textbf{(2)} Second, history is employed to estimate the policies of other agents. Notably, even in fully observable environments where the state is known, history remains essential for decision-making if the policies of other players are not directly accessible. This necessity for opponent modeling will be further discussed in Section~\ref{Section: Opponent Modeling}.

The \textit{policy learning rule} dictates how the agent updates its policy parameters, $\theta^i$ (when $\pi^i$ is a parameterized function $\pi^i_{\theta^i}$), to maximize this objective. The general learning rule involves an iterative update mechanism, typically derived from gradient-based optimization:
\begin{equation*}
\theta_{k+1}^i \leftarrow \theta_k^i + \alpha \nabla_{\theta^i} J^i(\boldsymbol{\pi}),
\label{eq:learning_rule}
\end{equation*}
where $k$ is the iteration index and $\alpha$ is the learning rate. We can abstract this update step for agent $i$ as a \textit{learning dynamics} $\mathcal{L}^i$, such that:
$$
\theta_{k+1}^i = \mathcal{L}^i(\theta_k).
$$
This is a \textit{vanilla} learning dynamics in an independent learning style, which considers its own parameters only.
The gradient $\nabla_{\theta^i} J^i(\boldsymbol{\pi})$ is estimated using various techniques such as policy gradient methods (e.g., REINFORCE~\citep{williams1992simple, sutton1999policy}, Actor-Critic~\citep{konda1999actor}) or Q-learning variants, where the form of the gradient estimate strongly depends on the specific MARL algorithm employed.

\subsection{Non-Stationarity}

The fundamental challenge that distinguishes multi-agent learning from single-agent learning is \textbf{non-stationarity}~\citep{hernandez2017survey}. In a single-agent system, the environment is governed by a stationary MDP. In contrast, in a multi-agent system, the effective environmental dynamics from the perspective of an ego agent $i$ are co-determined by the changing policies $\boldsymbol{\pi}^{-i} = (\pi^j)_{j \ne i}$ of all other agents. We can denote the subjective view of the environment and opponents from agent $i$ as the \textit{perspective environment} $\mathcal{G}^i$:
$$
\mathcal{G}^i = \langle \mathcal{G}, \boldsymbol{\pi}, \{\mathcal{L}^j \}_{j\in N} \rangle,
$$
and this perspective environment is non-stationary, i.e., even if the objective environment $\mathcal{G}$ and the ego agent's policy remain static, the ego agent's expected discounted return $J^i$ may change, due to the updates of other agents' policies.

Specifically, consider the vanilla learning rule as described above (i.e., the independent learning style), when any opponent $j \ne i$ updates its policy parameters via its learning rule $\mathcal{L}^j$ (i.e., $\theta^j_{k+1} = \mathcal{L}^j\left( \theta^j_k \right)$, this results in a change to its policy $\pi^j_k \to \pi^j_{k+1}$ (where $\pi^j_k$ is shorthand for $\pi^j_{\theta^j_k}$). This, in turn, changes the joint policy of all opponents, $\boldsymbol{\pi}^{-i}_k \to \boldsymbol{\pi}^{-i}_{k+1}$.
Consequently, even if agent $i$'s policy $\pi^i$ remains static, its expected discounted return $J^i$ is no longer stationary:
$$
J^i(\pi^i, \boldsymbol{\pi}^{-i}_k) \ne J^i(\pi^i, \boldsymbol{\pi}^{-i}_{k+1}).
$$
This instability invalidates the convergence guarantees of standard single-agent RL algorithms.


\subsection{Opponent Modeling}
\label{Section: Opponent Modeling}

To succeed in this non-stationary landscape, an agent $i$ must adapt its policy to the changing policies of its opponents. The goal shifts from finding a single optimal policy to dynamically computing a \textit{best response (BR)} to the current policies of the other agents. 

Calculating a BR presupposes that the agent possesses an estimate of its opponents' policies, $\hat{\boldsymbol{\pi}}^{-i}$. This inevitably leads to the need for \textit{opponent modeling}~\citep{he2016opponent}: the process of defining a function $m$ that maps available data (e.g., interaction history) to an estimate of the opponent policies:
$$
\hat{\boldsymbol{\pi}}^{-i} = m(h_t).
$$
To best respond to a hypothesized opponent policy, $\hat{\pi}^j$, the opponent modeling component $m(h_t)$ serves as a direct input to the agent's response policy. For example, a history-dependent best-response policy is defined as $\pi^i_{\text{br}}: \Pi^{-i} \times \mathbb{H} \to \Pi^i$. With this policy, player $i$ takes an action $a^i_t$ at timestep $t$ as:
$$
a^i_t \sim \pi^i_{\text{br}}(\cdot \mid \hat{\boldsymbol{\pi}}^{-i}, h_t) 
= \pi^i_{\text{br}}(\cdot \mid m(h_t), h_t) .
$$
It is important to note that inferring an opponent’s policy from the interaction history $h_t$ presents two primary challenges.
\textbf{(1)} First, the challenge of \textbf{\textit{policy drift}}: all policies are constantly updating, meaning the interaction data is not sampled from a stable joint policy distribution.
\textbf{(2)} Second, all agents are simultaneously trying to model each other, which introduces a complex hierarchy of beliefs~\citep{gmytrasiewicz2005framework, rabinowitz2018machine}. For example, agent $i$ must not only model what agent $j$ will do, but also model what agent $j$ guesses agent $i$ will do, and so on, leading to an \textbf{\textit{arbitrary-level recursive reasoning challenge}}~\citep{gmytrasiewicz2005framework}.










\subsection{Program Equilibrium and Provability Logic}

In canonical game theory, players select actions or strategies directly. The concept of \textit{Program Equilibrium}~\citep{tennenholtz2004program} departs from this by assuming that players delegate decision-making to computer programs. Crucially, these programs are capable of reading and conditioning their execution on the \textit{source code} of their opponent's program.

Formally, let $\mathbb{A}^i$ denote the action space for player $i$. Let $\Pi_{\text{prog}}$ denote the set of all valid computer programs in a given Turing-complete language. A policy for player $i$ is a program $\pi^i \in \Pi_{\text{prog}}$. To be rigorous, we explicitly distinguish between the program's execution logic (semantics) and its representation (syntax):
\begin{itemize}
    \item \textbf{Semantics ($\pi^i$):} We identify the policy $\pi^i$ directly with the computable function it implements upon execution.
    \item \textbf{Syntax ($\ulcorner \pi^i \urcorner$):} The G\"odel number or string representation of the program, where $\ulcorner \cdot \urcorner: \Pi_{\text{prog}} \to \{0,1\}^*$ is the encoding mapping.
\end{itemize}
Since $\pi^i$ acts as a function that takes the opponent's source code as input and maps it to a probability distribution over actions, the interaction is formally defined by:
$$
    \pi^i: \{0,1\}^* \to \Delta(\mathbb{A}^i)
$$
Specifically, in a game, player $i$ executes $\pi^i$ on input $\ulcorner \boldsymbol{\pi}^{-i} \urcorner$ to determine their action $a^i \sim \pi^i(\ulcorner \boldsymbol{\pi}^{-i} \urcorner)$. This meta-game structure enables outcomes that are inaccessible in standard Nash equilibria. For example, in the one-shot Prisoner's Dilemma, mutual cooperation becomes sustainable if players submit programs that verify whether the opponent's source code is functionally equivalent to their own (or satisfies a specific cooperative predicate) before cooperating.


A rigorous treatment of this self-referential structure, where program $A$ analyzes program $B$ analyzing program $A$, requires the machinery of \textit{Provability Logic}. In this context, agents are modeled as searching for proofs within a formal system (e.g., Peano Arithmetic). Let $\square P$ denote that the statement $P$ is provable. A canonical example is \texttt{FairBot}~\citep{barasz2014robust}, an agent that cooperates if and only if it can prove that the opponent cooperates against it. Formally, let $\pi_{\text{FB}}$ denote the \texttt{FairBot} program. It seeks a proof that the opponent's execution on its own source code yields cooperation:
$$
    \pi_{\text{FB}}(\ulcorner \boldsymbol{\pi}^{-i} \urcorner) = \text{``Cooperation''} \iff \square (\boldsymbol{\pi}^{-i}(\ulcorner \pi_{\text{FB}} \urcorner) = \text{``Cooperation''})
$$

While this creates a potential circular dependency, consistency is resolved via L\"ob's Theorem from mathematical logic, which states that for any formula $P$:
$$
    \square (\square P \to P) \to \square P
$$
This theorem implies that if an agent can prove ``if my cooperation is provable, then it is true,'' it can indeed prove its own cooperation. This mechanism creates a \textit{modal fixed point}, allowing mutual cooperation to be the provable outcome between two logical agents. This effectively establishes a \textit{\textbf{Folk Theorem}} for program games, implying that any feasible and individually rational payoff profile can be sustained in a one-shot setting through conditional code commitment, achieving outcomes that are typically restricted to repeated interactions.
An illustrative analysis of the program equilibrium within the Prisoner's Dilemma is provided in Appendix~\ref{Section: Program Equilibrium in the Prisoner's Dilemma}.

Despite its theoretical power, canonical Program Equilibrium faces significant practical hurdles. Formal verification of arbitrary Turing-complete policies is computationally undecidable (by Rice's Theorem) and brittle to trivial syntactic variations. Our work can be viewed as a relaxation of this rigorous requirement. By utilizing LLMs as \textit{approximate interpreters}, we operationalize a practical form of program equilibrium. Here, policies condition on the inferred \textit{semantics} of the opponent's code rather than requiring formal deductive proofs, thereby bridging the gap between theoretical game-theoretic commitments and modern multi-agent learning.

\section{Method}

This section presents our methodology. We first articulate the ``representational bottleneck'' in deep MARL, which inhibits conditioning on opponent policies. We then introduce our core solution: a paradigm shift to programmatic policies. This shift enables us to lift optimization from policy space to operator space, where the objective becomes learning a point-wise best-response operator, $\varphi^i$, that maps opponent policy-code to ego-agent policy-code. Finally, we introduce the Programmatic Iterated Best Response (PIBR) algorithm, a learning framework designed to optimize this operator.

\subsection{The Representational Bottleneck of Deep Reinforcement Learning}

To address the two mentioned challenges in opponent modeling (the challenges of policy drift and arbitrary-level recursive reasoning), it is helpful to introduce the communication perspective.
If an opponent $j$ directly transmits its policy $\pi^j$ via a signal $\sigma^j = \phi(\pi^j)$, the modeling task for agent $i$ becomes:
$$\hat{\pi}^j = m(\sigma^j) = m(\phi(\pi^j)),$$
where $\phi$ is $j$'s policy encoder and $m$ is $i$'s opponent modeling component which serves as a policy decoder.
Notice that the opponent modeling for agents no longer needs to be inferred from historical data, but is instead communicated through signals between the agents.

Agent $i$ takes an action sampled from its policy $\pi^i_{\text{br}}$, which is conditional on its estimation of opponents' policies:
$$
a^i_t 
\sim \pi^i_{\text{br}}\left( \cdot \mid h_t, \hat{\pi}^j \right) 
= \pi^i_{\text{br}} \left( \cdot \mid h_t, m(\sigma^{-i}) \right)
= \pi^i_{\text{br}} \left( \cdot \mid h_t, m(\phi(\pi^j)) \right).
$$
The signal, conceptualized as a latent variable, must satisfy two critical and often competing constraints:
\textbf{(1)} it must encapsulate and preserve maximal information pertaining to the policy of agent $j$;
\textbf{(2)} it must be encoded in a representational format that is simultaneously tractable and parsable by the decoder or actor-network of agent $i$. 
Consequently, the selection of an appropriate signal space is a pivotal design consideration.
In canonical MARL frameworks, this signal is often implicitly defined as the shared interaction history among agents.


In this work, we consider an extreme case in which $j$'s policy encoder $\phi$ is a source code extractor and $i$'s policy decoder $m$ is a identity function (i.e., $\phi(\pi^j) = \ulcorner \pi^j \urcorner$ and $m = \mathcal{I}$), yielding a perfect observation $\hat{\pi}^j = \ulcorner \pi^j \urcorner$.
We denote this setting as the \textit{\textbf{unprocessed policy-communication setting}}.
This setting is possible and can be justified in fully cooperative environments where there is no incentive for deception, or in zero-sum games under self-play~\citep{silver2017mastering} where the multi-agent system serves as an equilibrium solver.

A key advantage of this setting is that it inherently avoids any information loss that might arise from the communication process and obviates the need for learning a dedicated encoding mechanism. 
This setting also \textbf{\textit{avoids the policy drift and recursive reasoning challenges in opponent modeling}}, since agents can accurately obtain each other's policies through communication before acting, rather than having to guess.

The critical drawback, however, is that this approach places a substantial, perhaps prohibitive, parsing burden on the ego agent, which must now interpret the source code of policy directly:
$$
a^i_t \sim \pi^i\left( \cdot \mid h_t, \ulcorner \boldsymbol{\pi}^{-i} \urcorner \right).
$$
This means that $\pi^i$ will take other $\ulcorner \boldsymbol{\pi}^{-i} \urcorner$ as input. If all agents' policies are implemented by neural networks as in MARL, this means that $\pi^i$'s neural network must parse the entire neural network of $\pi^j$ (i.e., its complete parameter vector $\theta^j$) as input.
This leads to two insurmountable obstacles:
\textbf{(1) \textit{Curse of Dimensionality.}} A modern policy network can have millions or billions of parameters. The requirement for $\pi^i$'s input layer to accommodate and process this magnitude of parameters is computationally prohibitive. It also creates a recursive loop: $\pi^i$ must grow to parse $\pi^j$, and $\pi^j$ must in turn grow to parse the now-larger $\pi^i$.
\textbf{(2) \textit{Curse of Representation.}} This is not just a problem of dimensionality, but of representation. The parameter vector $\theta^j$ is a brittle and non-unique representation of the policy function. A neural network's function is invariant to permutations of its hidden nodes; different random initializations lead to functionally similar policies with drastically different parameter vectors. $\pi^i$ would need to learn a complex, permutation-invariant representation over this unstructured, high-dimensional vector, which is a task for which there is currently no viable solution.


\subsection{A Paradigm Shift: Programmatic Policies by LLMs}


We posit that, while DRL faces challenges in realizing the unprocessed policy-communication Setting, the use of recently flourishing LLMs to generate code as policies does possess the capacity to handle such policy-input.
Its objective is to supplant the opaque, ``black box'' nature of conventional DRL with policy representations that are structured, interpretable, and more amenable to formal verification.

This paradigm shift is facilitated by the advent of LLMs. LLMs are inherently designed to process structured text and source code. Modern architectures are not only proficient code generators but also feature extensive context windows, capable of managing inputs equivalent to hundreds of thousands of tokens.

This framework directly addresses the aforementioned intractability challenges:
\textbf{(1) \textit{Solving Dimensionality}}: While a neural network policy may require millions or billions of parameters to capture complex behaviors, its programmatic equivalent offers a highly compressed, semantically dense representation. Modern LLMs, with their extensive context windows, are capable of processing and reasoning over such complex programmatic structures, thus effectively managing the high intrinsic dimensionality of the policy space.
\textbf{(2) \textit{Solving Representation}}: Unlike the unstructured parameter vectors of a neural network, source code is a semantically rich and structured artifact. This representation aligns precisely with the data modalities that LLMs are trained to comprehend, analyze, and generate, enabling high-level reasoning about, and manipulation of, the policy itself.

\subsection{Refined Formalization: Meta-Programming and Mathematical Operators}

We leverage this new paradigm to reformalize the best-response computation.
Mathematically, this process is most precisely described using the terminology of functional analysis. The space of all policies $\Pi$ constitutes a function space.
An \textit{operator} maps one function space to another; a canonical example is the differential operator $D$, which maps a function $f$ to its derivative $f'$, i.e., $D(f) = f'$.

Following this definition, our proposed process (denoted as $\varphi^i$) is formally a \textbf{best-response operator}. It constitutes a mapping $\varphi^i: \{0,1\}^* \to \{0,1\}^*$ that takes an opponent policy function as input and yields a new ego-agent policy function as output. This operator, $\varphi^i$, thus computes the ``best response'' to an input policy function. 

From an implementation perspective, this abstract operator $\varphi^i$ is realized through a \textbf{meta-programming} interpreter. Meta-programming is formally defined as the act of a program treating other programs as its data; in this context, it manifests as a code-input, code-output process.
This operation is implemented by querying a LLM. Agent $i$ provides its opponent's policy source code $\boldsymbol{\pi}^{-i}$ as input (i.e., as part of the prompt), and the LLM, embodying the operator $\varphi$, generates the best-response policy source code $\pi^i$ as output.
Policy adaptation is thus rendered as an interpretable meta-programming task.

Formally, given opponents' policies $\pi^{-i}$, the ego agent will take an action as follows:
\begin{equation*}
\begin{aligned}
& \ulcorner \boldsymbol{\pi}^{-i} \urcorner = m\left(\phi\left(\boldsymbol{\pi}^{-i}\right)\right), \\
& \ulcorner \pi^i \urcorner = \varphi^i(\ulcorner \boldsymbol{\pi}^{-i} \urcorner \mid \text{prompt}), \\
& \pi^i = \text{CodeInterpreter}(\ulcorner \pi^i \urcorner), \\
& a^i\sim \pi^i,
\end{aligned}
\end{equation*}
where $\varphi^i$ is implemented by an LLM in our method. 
Note that the policy of ego agent $\pi^i$ does not take opponents' source code $\ulcorner \boldsymbol{\pi}^{-i} \urcorner$ as inputs, but it still has the dependency since the generation of $\pi^i$ is dependent on $\ulcorner \boldsymbol{\pi}^{-i} \urcorner$.

While any generated $\pi^i$ constitutes a valid response, it is not inherently a best response; we therefore seek the optimal $\pi^i$ by optimizing the operator $\varphi^i$ through prompt refinement. 
Crucially, since we optimize the operator specifically for a fixed input $\boldsymbol{\pi}^{-i}$, this constitutes a \textbf{point-wise} optimization of the best-response mapping.

\subsection{Learning}

We optimize the LLM best-response operators $\varphi$ in prompt space. This optimization is driven by structured feedback losses and can be automated using gradient-based methods adapted for text. A notable framework in this domain is \texttt{TextGrad}~\citep{yuksekgonul2024textgrad},which enables automatic differentiation through text-based models by conceptualizing the text generation process as a computational graph. This allows for the definition of textual feedback and the backpropagation of gradients to update the LLM's guiding prompts or parameters. Once we rigorously define the loss functions, the optimization of the operator $\varphi^i$ can proceed automatically.

\paragraph{Textual Loss Functions.}
The feedback provided to the operator $\varphi^i$ is structured around two primary loss signals:
\textbf{(1)} \textit{Unit Test Feedback.} A significant challenge in this paradigm is ensuring the syntactic and runtime correctness of the generated policy code. To mitigate this, we employ an automated unit testing phase. The ego policy code $\pi^i = \text{CodeInterpreter}(\ulcorner \pi^i \urcorner)$ is instantiated and executed within a sandboxed environment. If the code produces any runtime exceptions (e.g., syntax errors, out-of-bounds access) or fails predefined assertions, the resulting error trace and stack information are captured. This error data serves as a direct, high-priority feedback signal, prompting the LLM operator $\varphi^i$ to debug and revise the code, akin to automated debugging workflows.
\textbf{(2)} \textit{Utility Feedback.} The central objective is to ensure the generated policy $\pi^i$ constitutes a best response to the opponents' policies $\boldsymbol{\pi}^{-i}$. This is evaluated via utility feedback. The policy $\pi^i$ is executed against $\boldsymbol{\pi}^{-i}$ in the game environment for several episodes. The complete episode trajectory, including all states, joint actions, and rewards for both agents $(s_t, a^i_t, \boldsymbol{a}^{-i}_t, r^i_t, \boldsymbol{r}^{-i}_t)_{t=0}^T$, is collected. This detailed trajectory information is provided as feedback to the operator $\varphi^i$. This allows the operator to assess the performance of its generated policy and refine its meta-programming logic to maximize the expected utility $J^i(\pi^i, \boldsymbol{\pi}^{-i})$.




The overall algorithm follows an iterative best-response dynamic~\citep{robinson1951iterative}, but lifted to the operator space. The computation of the best response for each agent is not an atomic operation but an inner optimization loop guided by feedback.
This two-agent version of this process is formally described in Algorithm~\ref{alg:os_fp}. 


\begin{algorithm}[t]
\caption{Programmatic Iterated Best Response (PIBR)}
\label{alg:os_fp}
\SetKwProg{Proc}{Procedure}{}{}
\SetKwComment{Comment}{// }{}
\DontPrintSemicolon

\KwIn{Game $G$, Agents $A = \{A^0, A^1\}$, Outer loops $K$, Inner loops $T$}
\KwOut{Final policy profile $\boldsymbol{\pi}^* = (\pi^{0, *}, \pi^{1, *})$}

\tcp{Initialize policies and history}
$\pi^0 \gets \text{InitialPolicy}()$; $\pi^1 \gets \text{InitialPolicy}()$\;
$\mathcal{H} \gets \emptyset$ \Comment{History of (profile, social welfare)}
$i \gets 0$ \Comment{Agent index}

\tcp{Main training phase (Outer Loop)}
\For{$k = 1$ \KwTo $K$}{
    $\pi^{-i} \gets \pi^{1-i}$\;
    
    \tcp{Compute BR via operator optimization (Inner Loop)}
    $\pi^{i} \gets \text{ComputeBestResponse}(A^i, G, \pi^{-i}, T)$\;
    
    $\boldsymbol{\pi}_k \gets (\pi^0, \pi^1)$\;
    $J_{sw}(\boldsymbol{\pi}_k) \gets \text{EvaluateSocialWelfare}(G, \boldsymbol{\pi}_k)$\;
    $\mathcal{H} \gets \mathcal{H} \cup \{ (\boldsymbol{\pi}_k, J_{sw}(\boldsymbol{\pi}_k)) \}$\;
    
    $i \gets 1 - i$ \Comment{Alternate agent}
}

\tcp{Select best profile from history}
$(\boldsymbol{\pi}^*, J^*) \gets \arg\max_{(\boldsymbol{\pi}, J) \in \mathcal{H}} J$\;
\Return $\boldsymbol{\pi}^*$\;

\vspace{0.5em}
\hrule
\vspace{0.5em}

\Proc{\text{ComputeBestResponse}($A^i, G, \pi^{-i}, T$)}{
    $\varphi^i \gets A^i.\text{operator}$ \Comment{Get optimizable operator}
    $\text{Opt} \gets \text{TextGradOptimizer}(\varphi^i)$\;
    
    \For{$t = 1$ \KwTo $T$}{
        $\text{Opt.zero\_grad}()$\;
        
        $\ulcorner \pi^i_t \urcorner = \varphi^i\left(\ulcorner \pi^{-i} \urcorner \mid \text{prompt}(A^i, G, t, T)\right)$ \Comment{Forward pass}
        
        \Comment{Compute composite loss}
        $(\pi^i_t , \mathcal{L}_{\text{test}}) \gets \text{ComputeUnitTestLoss}(\ulcorner \pi^i_t \urcorner)$\;
        $\mathcal{L}_{\text{utility}} \gets -J(G, \pi^i_t, \pi^{-i})$
        $\mathcal{L} \gets \mathcal{L}_{\text{test}} + \mathcal{L}_{\text{utility}}$\;
        
        $\mathcal{L}.\text{backward}()$\;
        $\text{Opt.step}()$\;
    }
    \Return $\pi^i_t$
}
\end{algorithm}


\section{Experiments}

All experiments were powered by OpenAI’s \texttt{o4-mini} model~\citep{openai_o4mini_2025}, as it provides a particularly effective balance between cost efficiency and code generation performance.
The complete codebase, experimental results, and execution logs will be released as open source after the paper is accepted.

\subsection{Tasks}
\label{Section: Tasks}

\paragraph{Matrix Games.} 
We consider three representative coordination matrix games that vary in their payoff structures and coordination difficulty. In each game, two agents simultaneously choose an action (row and column respectively), and both receive the same payoff corresponding to the selected matrix entry. 
\textbf{(a)} The \texttt{Vanilla Coordination Game} is a simple setting with three pure coordination equilibria of different rewards, testing whether agents can align on the optimal coordination point. 
\textbf{(b)} The \texttt{Climbing Game}~\citep{claus1998dynamics} introduces strong miscoordination penalties and uneven reward gradients, making it a challenging variant where naive exploration can easily lead to suboptimal equilibria. 
Finally, \textbf{(c)} the \texttt{Penalty Game}~\citep{claus1998dynamics} includes negative diagonal payoffs ($p<0$) that discourage coordination on certain actions, requiring agents to learn to avoid detrimental equilibria while still maximizing joint payoff. Together, these games provide a controlled yet diverse testbed for studying coordination behaviors in multi-agent learning.

\begin{figure}[ht!]
\centering
\setlength{\tabcolsep}{1pt} 
\renewcommand{\arraystretch}{1.2}

\begin{tabular}{ccc}

\begin{minipage}[t]{0.33\textwidth}
\centering
\[
\begin{array}{c|ccc}
\multicolumn{1}{c|}{\quad} &     & a^j &     \\
\cline{1-4}
    & 2 & 0 & 0\\
a^i & 0 & 1 & 0\\
    & 0 & 0 & 3
\end{array}
\]
(a) Vanilla Coordination Game
\end{minipage}
&
\begin{minipage}[t]{0.33\textwidth}
\centering
\[
\begin{array}{c|ccc}
\multicolumn{1}{c|}{\quad} &     & a^j &     \\
\cline{1-4}
    & 11 & -30 & 0\\
a^i & -30 & 7 & 0\\
    & 0 & 6 & 5
\end{array}
\]
(b) Climbing Game
\end{minipage}
&
\begin{minipage}[t]{0.33\textwidth}
\centering
\[
\begin{array}{c|ccc}
\multicolumn{1}{c|}{\quad} &     & a^j &     \\
\cline{1-4}
    & p & 0 & 10\\ 
a^i & 0 & 2 & 0\\
    & 10 & 0 & p
\end{array}
\]
(c) Penalty Game ($p<0$)
\end{minipage}
\end{tabular}

\caption{Three $3 \times 3$ coordination matrix games used in our experiments. Each game defines payoffs for two agents (row agent's action $a^i \in \{0,1,2\}$, column agent's action $a^j \in \{0,1,2\}$): after they simultaneously choose a row and a column action, both receive the utility indicated by the selected matrix entry.}
\label{Fig: matrix-games}
\end{figure}

\paragraph{Level-Based Foraging.} Level-Based Foraging (LBF)~\citep{christianos2020shared} is a grid-world environment for studying multi-agent coordination. 
Agents are to collect food items to gain rewards.
Each agent can take one of six actions: \texttt{[stay\_idle, move\_up, move\_down, move\_left, move\_right, load\_food]}. 
In the original environment, food items can only be collected when the sum of the levels of agents performing the \texttt{load} action adjacent to the food meets or exceeds the food’s level, making it a mixed-motive task.
\footnote{E.g., collecting Level 1 foods in the grid-world is competitive among Level 1 agents, but they must cooperate to collect Level 2 foods.}
We consider a fully cooperative variant here: each food is assigned a level such that it always requires both agents to perform the load action simultaneously to be collected. 
Rewards are shared between agents proportionally to their skill levels and normalized.
An episode ends when all foods are collected or the maximum length is reached. 
We further modify the environment by introducing a terminal penalty that is inversely proportional to the episode length, making the improvement of food collection efficiency the optimization objective for the agents.

\subsection{Results}

\begin{figure*}[t]
    \centering
    \captionsetup[subfigure]{aboveskip=2pt, skip=2pt}
    
    \begin{subfigure}[b]{0.49\textwidth}
        \centering
        \includegraphics[width=1\textwidth]{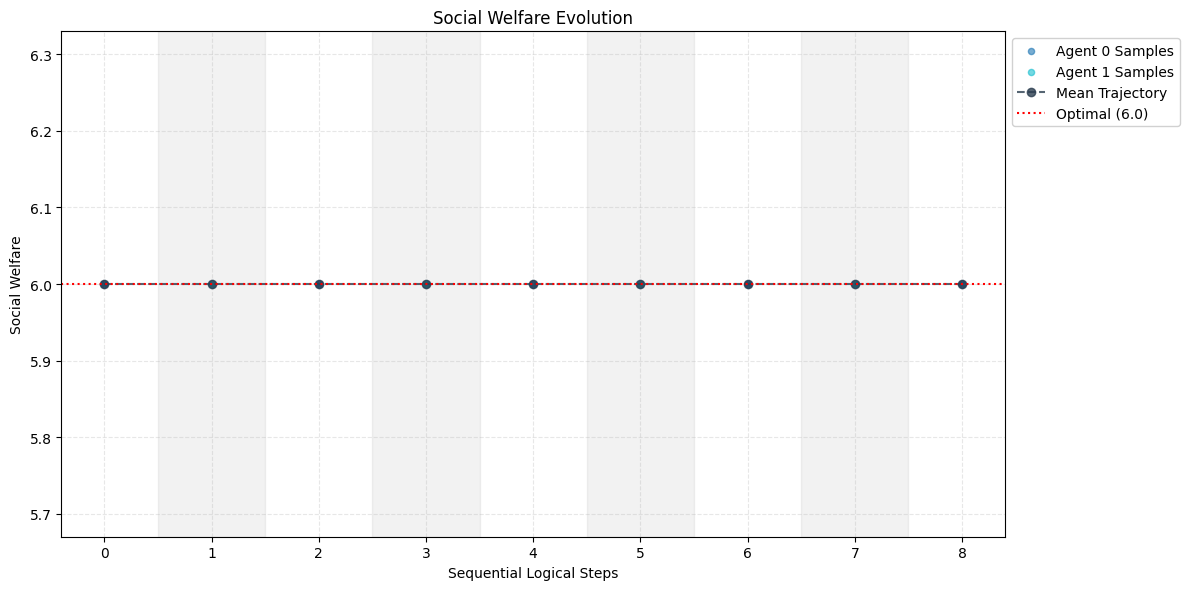}
        \caption{Vanilla Coordination Game}
        \label{fig:vanilla}
    \end{subfigure}%
    \begin{subfigure}[b]{0.49\textwidth}
        \centering
        \includegraphics[width=1\textwidth]{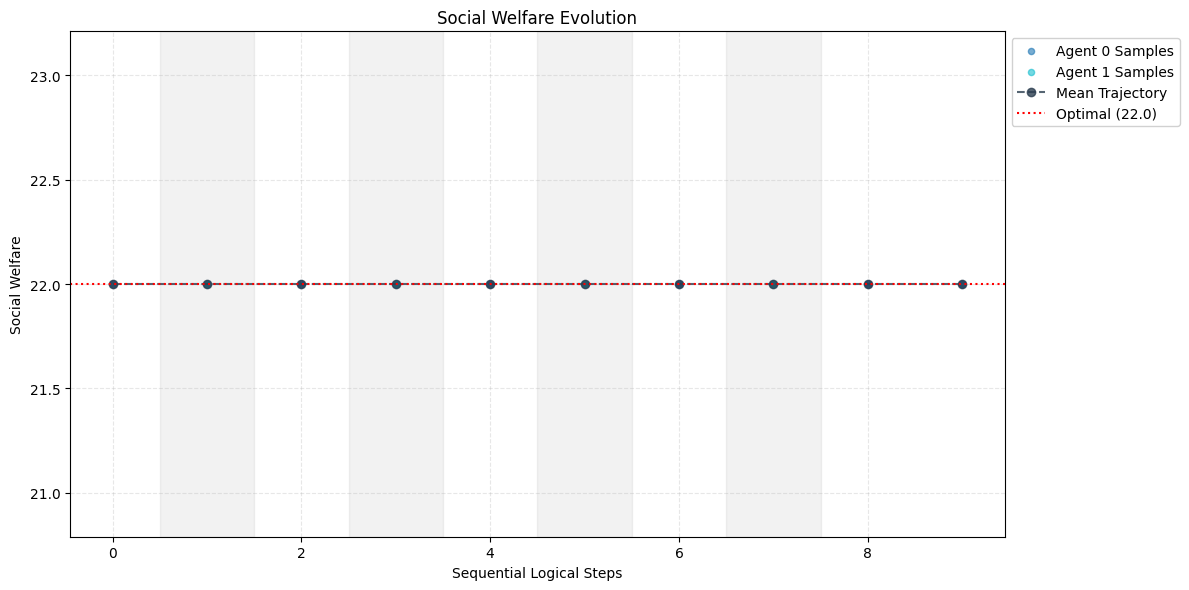}
        \caption{Climbing Game}
        \label{fig:climbing}
    \end{subfigure}

    \vspace{0.2em} 

    \begin{subfigure}[b]{0.49\textwidth}
        \centering
        \includegraphics[width=1\textwidth]{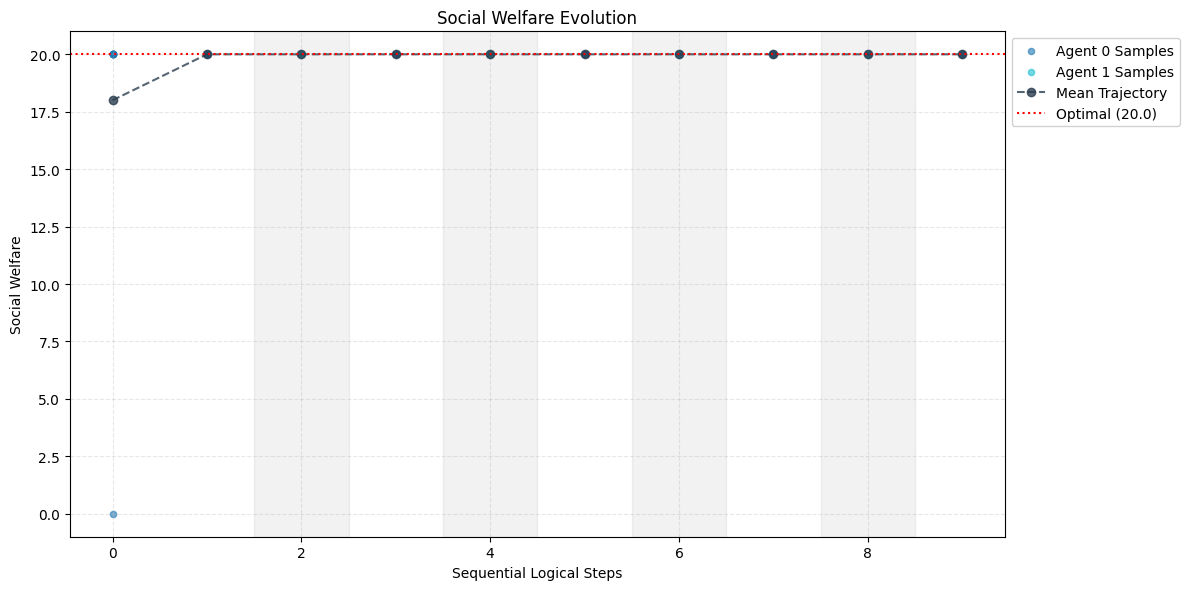}
        \caption{Penalty Game ($p=-2$)}
        \label{fig:penalty}
    \end{subfigure}%
    \begin{subfigure}[b]{0.49\textwidth}
        \centering
        \includegraphics[width=1\textwidth]{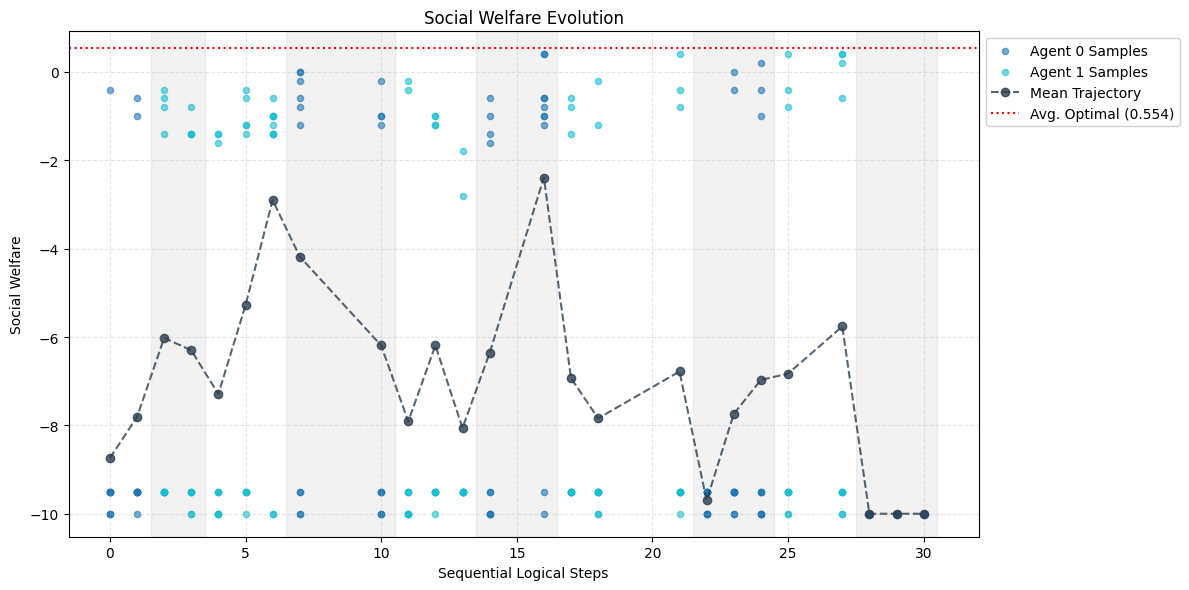}
        \caption{Cooperative Foraging}
        \label{fig:foraging}
    \end{subfigure}

    \caption{Experimental results.}
    \label{fig:four_experiments}
\end{figure*}

The experimental results, summarized in Figure \ref{fig:four_experiments}, demonstrate the effectiveness of PIBR across distinct game types.
In these plots, the horizontal axis represents the sequential optimization steps, where each step corresponds to a policy update step driven by textual gradients. The vertical axis denotes the Social Welfare, calculated as the sum of cumulative rewards obtained by the agents. 
For every optimization step, we sample multiple episodes to evaluate the current policy profile; these individual episodes are visualized as scattered points, while the dashed line traces the mean reward trajectory across optimization steps.
The absence of data points at certain optimization steps indicates that the generated policy code contained syntax or runtime errors, failing to execute in the environment.
\textbf{(1) Results of Coordination Games:} across all three coordination tasks, the agents exhibit rapid and robust convergence to the global optima. In \texttt{Vanilla Coordination} and \texttt{Climbing Game}, the social welfare instantaneously stabilizes at the optimal values of $6.0$ and $22.0$, respectively. In the \texttt{Penalty Game}, despite the risk of substantial penalties, the agents successfully coordinate on the optimal equilibrium within a single update step.
\textbf{(2) Results of Cooperative Foraging:} In this complex grid-world task, while individual samples occasionally approach the empirical optimum ($\approx 0.554$). 
The mean trajectory exhibits significant variance and instability. The performance drop in later stages indicates the challenge of maintaining consistent coordination in complicated tasks.
Following Line $11$ of Algorithm~\ref{alg:os_fp}, the procedure returns the agents' strategies from the $16$-th optimization step, since it yields the highest social welfare.
We recognize that the current results are not yet fully stabilized, and further improvements to the method's robustness are under active development.

    
    

\section{Conclusion}


In this work, we address the challenge of non-stationarity in solving Markov games by proposing a paradigm shift from opaque neural policies to interpretable programmatic policies, thereby overcoming the ``representational bottleneck'' that hinders direct conditioning on opponent strategies. We reformulate the learning problem by lifting optimization from the policy space to the operator space, employing LLMs as point-wise best-response operators that dynamically synthesize policy code based on the opponent's source code. We operationalize this via the Programmatic Iterated Best Response algorithm, which optimizes the meta-programming logic through textual gradients derived from both game utility and runtime unit test feedback. The proposed method seeks program equilibria rather than black-box Nash equilibria, facilitating a broader scope of cooperation. Our experiments in coordination matrix games and Level-Based Foraging demonstrate that PIBR effectively bridges the gap between theoretical Program Equilibrium and modern learning, enabling the discovery of robust, coordinated strategies through the exchange and analysis of executable code.


\bibliography{arxiv}
\bibliographystyle{arxiv}

\appendix

\newpage
\addcontentsline{toc}{section}{Appendix}
\part{Appendix}
\parttoc

\section{Related Works}
\label{Section: Related Works}

\subsection{Opponent Modeling}


A large body of MARL research can be categorized by how it approaches this modeling function $m$.
\textbf{(1)} \textit{Implicit Modeling.}
A dominant branch of research, particularly within the CTDE framework, models opponents implicitly. In methods like Value-Decomposition Networks (VDN)~\citep{sunehag2017value}, QMIX~\citep{rashid2020monotonic}, and MADDPG~\citep{lowe2017multi}, a centralized critic is provided with global state and action information from all agents during training. 
This critic implicitly learns the interaction dynamics, conveys information by \textit{parameter sharing}, and uses this knowledge to guide the decentralized actors.
The resulting actors themselves do not have an explicit mechanism at execution time to adapt to novel opponent policies not seen during training.
\textbf{(2)} \textit{Explicit Modeling.}
An alternative approach attempts to learn $\hat{\pi}^{-i}$ explicitly. Works like \citet{he2016opponent} design neural architectures aimed explicitly at predicting the opponent's behavior or encoding it into a latent space. While theoretically more adaptive, these methods face the challenge of learning a high-dimensional mapping from history to the policy, which is also a high-dimensional object.

There is also some work that goes a step further by being aware of the other agent's learning dynamics to better predict the ego agent's expected value~\citep{foerster2017learning}, or even leverages mechanisms to alter the other agent's behavior~\citep{yang2020learning, lin2023information}.

\subsection{Commitment}

A central challenge in committing on other agents' source code is ensuring credible, robust, and non-exploitative cooperation when agents can read, simulate, or reason about each other's programs. Classical program equilibrium~\citep{tennenholtz2004program} shows that full cooperation in one-shot games is achievable when programs condition on opponents' code, but these early constructions are brittle and depend on exact code matching. Logical approaches using L{\"o}b's theorem demonstrate that agents can instead cooperate when they can prove that the opponent will reciprocate, yielding robust program equilibria without requiring identical code~\citep{lavictoire2014program,barasz2014robust}, though still vulnerable to issues of exploitation and representational fragility. Follow-up work extends these results to resource-bounded agents via parametric bounded L{\"o}b reasoning~\citep{critch2016parametric}. When agents can simulate each other at a cost, cooperation can emerge when distrust is the only barrier, but simulation can also reduce welfare by enabling new threats~\citep{kovarik2023game}. Under incomplete information, conditional information disclosure provides commitment devices that sustain efficient outcomes~\citep{digiovanni2023commitment}.
Similarly, \citet{lin2025bayesian} frame the selection of these information structures as a bargaining game, enabling both sender and receiver agents to commit to signaling schemes that enforce cooperative equilibria. This parallels formal contracting frameworks in which agents write enforceable program-like contracts~\citep{peters2012definable}.
In practice, decentralized cryptographic mechanisms provide commitment infrastructure but introduce new security and integrity risks~\citep{sun2023cooperative}. Open-source game theory reveals counterintuitive or unstable outcomes even under full transparency~\citep{critch2022cooperative}, highlighting unresolved robustness challenges. Recent work on safe Pareto improvements addresses some of these risks by identifying commitment restrictions that are guaranteed not to harm any party~\citep{oesterheld2022safe}, while empirical studies of LLM agents in open-source games show both cooperative and deceptive behaviors, underscoring the need for practical safeguards when commitments rely on program inspection~\citep{sistlaevaluating}.

\subsection{Fictitious Play}

Fictitious play (FP)~\citep{robinson1951iterative,berger2007brown} is a classical game-theoretic learning dynamic in which each agent iteratively best-responds to the empirical frequency of others' past actions, often converging to a Nash equilibrium in certain classes of games~\citep{bachrach2025combining}. FP has been foundational in multi-agent learning and has been extended to richer settings where agents explicitly model or inspect their opponents' decision-making procedures. In the extreme case of open-source or program games~\citep{tennenholtz2004program}, each player submits an algorithm that can read the other's source code, enabling novel equilibria based on mutual transparency (e.g. two programs mutually cooperating in a one-shot dilemma)~\citep{fortnow2009program}. \citet{tennenholtz2004program}'s program equilibrium framework formalized how agents conditioning on opponents' code can achieve such outcomes beyond standard game-theoretic limits. These ideas are relevant to modern AI: LLM-based agents can simulate or infer an opponent's policy or reasoning process, effectively building a theory-of-mind. For instance, prompting an LLM to reason about an opponent's likely action via a social chain-of-thought markedly improves coordination in repeated games~\citep{akata2025playing}. The intersection of fictitious-play-style adaptation with LLMs' capacity for modeling other agents thus points to a promising avenue for multi-agent systems, where agents dynamically learn and adapt by reasoning about each other's strategies. 

\subsection{LLMs for Multi-Agent Tasks} 
Recently, due to the powerful reasoning capabilities and extensive intrinsic knowledge of LLMs, research on LLM-based multi-agent systems has progressed rapidly \citep{guo2024large,han2024llm}. This line of work can be broadly divided into two categories: social simulation and task collaboration. Social simulation research utilizes the natural language interaction capabilities of LLMs to analyze and calibrate emergent behaviors in different social interactions, benchmarking them against human society. \cite{park2023generative} develops the first sandbox environment driven by generative LLM agents to simulate a small-scale community of 25 agents and provides a user-interactive interface for engagement. CAMEL \citep{li2023camel} provides an scalable framework to support dialogue-intensive scenarios and keep consistency of agents during role playing, which is suitable for studying conversational behaviors. \cite{xu2023exploring} investigates emergent strategic behaviors—such as trust, confrontation, camouflage, and leadership—of LLMs in Werewolf, revealing the strong potential of LLMs in communication games.
Beyond simple interaction, \citet{li2025verbalized} explore strategic verbal communication, utilizing LLMs to operationalize Bayesian persuasion and influence recipients' beliefs through natural language.

In task collaboration, the objective is to develop algorithmic frameworks that coordinate multiple LLMs to complete complex tasks, including natural-language tasks and text-based games \citep{costarelli2024gamebench, schmidgall2025agent, qian2024chatdev}. MetaGPT \citep{hong2023metagpt} incorporates standardized operating procedures (SOPs) into multi-LLM workflows to improve coordination and the accuracy of intermediate processes, which decomposes software development task dinto multiple subtasks and roles. ProAgent \citep{zhang2024proagent} proposes a modular planning framework to facilitate coordination with arbitrary opponents, including belief, plan, memory, self-reflection, and action. GPTswarm \citep{zhuge2024gptswarm} represents multi-agent LLM systems as computational graphs, where each LLM agent is represented as a node and the communication among agents is represented as the edges. This flexible structure enables ensembling diverse LLM functions and optimizing workflow topologies. Multi-Agent Debate \cite{du2023improving} draws inspiration from debate and voting mechanisms in human society, using multiple rounds of reasoning and opinion exchange to improve accuracy on mathematical and QA problems.

However, these approaches often rely on natural-language dialogue to exchange information and intent, or treat LLM as executable policies with huge parameters. This leads to a lack of game-theoretic guarantees and commitment constraints in communication, which can introduce ambiguity and hallucinations into the messages. Furthermore, the large parameter space poses  severe challenges for policy optimization and coordination. Our approach effectively improves policy interpretability and reduces the search space by expressing the policy as executable code and communicating it explicitly.

\subsection{Code Policies by LLMs}
A growing body of work investigates using large language models to generate executable programs that function directly as control policies, offering interpretability and compositional structure absent in neural actors.
Code as Policies (CaP)~\citep{liang2023code} generates Python programs that compose motion primitives and conditional rules, producing explicit and adaptable robot-control policies.
Code-driven planning~\citep{aravindan2025code} formulates policy generation as program synthesis, producing loops and state-handling logic that act as executable decision policies in grid-world environments.
Hierarchical program generation~\citep{luo2024llm_hierachical} creates both high-level and low-level control code for construction-robot tasks, encoding sequencing and safety constraints as explicit policy logic.
RL-GPT~\citep{liu2024rlGPT} integrates reinforcement learning with LLM-generated program structures, allowing reward-driven refinement of high-level policy code.
Voyager~\citep{wangvoyager} continuously synthesizes and improves executable skill programs, treating each program as a reusable policy component for lifelong embodied learning.
Reflexion~\citep{shinn2023reflexion} iteratively updates code-like decision rules after execution failures, forming a self-improving program-structured policy.
Self-Refine~\citep{madaan2023self} applies critique-and-correct refinement to improve generated code, yielding progressively stronger programmatic decision rules.
Program-of-Thought prompting~\citep{chenprogram} generates executable intermediate programs that act as explicit policies in sequential decision-making tasks.
LLM-Planner~\citep{song2023llm} synthesizes executable planning scripts whose encoded procedures serve as interpretable multi-step policies.
SayCan~\citep{brohan2023can} outputs structured action sequences resembling policy scripts, using affordance constraints to guide task-level decision making.
DEPS~\citep{wang2023describe} produces decomposed control programs that function as hierarchical policies for navigation and manipulation tasks.
LLM-Guided Search (LLM-GS)~\citep{liusynthesizing} uses LLM priors to guide the synthesis of discrete programmatic RL policies, though DSL compliance remains a core limitation.
TaskMatrix.AI~\citep{Liang2024TaskMatrix} generates API-calling scripts that orchestrate multi-tool operations, serving as high-level executable policies.
LLM-driven strategy generation~\citep{willis2025will} translates LLM-produced strategies into executable code for iterated social-dilemma settings, enabling systematic evaluation of cooperative and adversarial policies.

\section{Program Equilibrium in the Prisoner's Dilemma}
\label{Section: Program Equilibrium in the Prisoner's Dilemma}

In this appendix, we explicitly derive the Program Equilibrium for the Prisoner's Dilemma (PD) using the formalism of Provability Logic. Recall that the standard PD payoffs satisfy $r_T > r_R > r_P > r_S$.

We analyze the stability of the \texttt{FairBot} program, denoted as $\pi_{\text{FB}}$. The source code of $\pi_{\text{FB}}$ is defined as follows:
\begin{equation}
    \pi_{\text{FB}}(\ulcorner \pi^{-i} \urcorner) = 
    \begin{cases} 
    \text{``Cooperation''} & \text{if } \mathcal{T} \vdash \pi^{-i}(\ulcorner \pi_{\text{FB}} \urcorner) = \text{``Cooperation''} \\
    \text{``Defect''} & \text{otherwise}
    \end{cases}
\end{equation}
where $\mathcal{T} \vdash \mathcal{P}$ denotes that the statement $\mathcal{P}$ is provable in the base theory $\mathcal{T}$ (e.g., Peano Arithmetic), denoted by $\square \mathcal{P}$ in modal logic.

To demonstrate that the profile $(\pi_{\text{FB}}, \pi_{\text{FB}})$ constitutes a Nash equilibrium, we examine both on-path feasibility and off-path stability.

\subsection{Feasibility: Mutual Cooperation via L\"ob's Theorem}
Consider the case where two agents both submit $\pi_{\text{FB}}$. Let $\mathcal{P}$ be the sentence asserting that \texttt{FairBot} cooperates against itself:
$$
\mathcal{P} := (\pi_{\text{FB}}(\ulcorner \pi_{\text{FB}} \urcorner) = \text{``Cooperation''})
$$
By the definition of the program, $\pi_{\text{FB}}$ cooperates if and only if it finds a proof that its opponent (which is itself) cooperates. Therefore, we have the fixed-point equivalence:
\begin{equation}
    \label{eq:fixed_point}
    \mathcal{T} \vdash (\mathcal{P} \leftrightarrow \square \mathcal{P})
\end{equation}
To prove that cooperation actually occurs ($\mathcal{P}$), we invoke \textbf{L\"ob's Theorem}, which states that for any sentence $A$, if $\mathcal{T} \vdash (\square A \to A)$, then $\mathcal{T} \vdash A$.

We verify the condition for L\"ob's Theorem as follows:
\begin{enumerate}
    \item From the program definition (\ref{eq:fixed_point}), we have the implication $\mathcal{T} \vdash (\square \mathcal{P} \to \mathcal{P})$.
    \textit{Intuitive explanation of the above implication:} Inside the formal system, if we hypothetically assume that a proof of cooperation exists ($\square \mathcal{P}$), the program's condition is satisfied, and it will output ``Cooperation'' ($\mathcal{P}$).
    \item Since $\mathcal{T} \vdash (\square \mathcal{P} \to \mathcal{P})$ holds, L\"ob's Theorem implies $\mathcal{T} \vdash \mathcal{P}$.
\end{enumerate}
Thus, the system proves that \texttt{FairBot} cooperates. Assuming the formal system $\mathcal{T}$ is sound (i.e., proven statements are true in reality), both programs output ``Cooperation'', yielding the payoff $r_R$.

\subsection{Stability: Robustness against Deviations}
We check for Nash equilibrium by considering a unilateral deviation. Suppose player $-i$ adheres to $\pi_{\text{FB}}$. We analyze whether player $i$ can gain by deviating to an arbitrary program $\pi' \in \Pi_{\text{prog}}$.

The equilibrium payoff is $u^i(\pi_{\text{FB}}, \pi_{\text{FB}}) = r_R$.

\textbf{Case 1: Cooperative Deviation.}
Suppose player $i$ submits a program $\pi'$ such that $\mathcal{T} \vdash \pi'(\ulcorner \pi_{\text{FB}} \urcorner) = \text{``Cooperation''}$.
Since $\pi_{\text{FB}}$ searches for precisely this proof, $\pi_{\text{FB}}$ will cooperate. The outcome is $(C, C)$, yielding payoff $r_R$. The deviation does not strictly increase the payoff.

\textbf{Case 2: Defecting Deviation.}
Suppose player $i$ submits a program $\pi'$ that actually defects against $\pi_{\text{FB}}$, i.e., $\pi'(\ulcorner \pi_{\text{FB}} \urcorner) = \text{``Defect''}$.
\texttt{FairBot} will only cooperate if it finds a proof of $\pi'$ cooperating:
$$
\square (\pi'(\ulcorner \pi_{\text{FB}} \urcorner) = \text{``Cooperation''})
$$
Assuming the base theory $\mathcal{T}$ is \textbf{consistent}, it cannot prove a false statement. Since $\pi'$ actually defects, the statement ``$\pi'$ cooperates'' is false. Therefore, no proof exists ($\neg \square (\text{``}\pi' \text{ cooperates''})$).
Consequently, the condition for $\pi_{\text{FB}}$ fails, and $\pi_{\text{FB}}$ outputs ``Defect''.

The resulting outcome is $(D, D)$, yielding payoff $r_P$. Since $r_R > r_P$ in the Prisoner's Dilemma, this deviation results in a strictly lower payoff.

$$
u^i(\pi_{\text{FB}}, \pi_{\text{FB}}) \ge u^i(\pi', \pi_{\text{FB}}) \quad \forall \pi'
$$
Thus, $(\pi_{\text{FB}}, \pi_{\text{FB}})$ is a Nash equilibrium.



\section{Generated Code-Policies}

\subsection{Generated Code-Policies for the Climbing Game}

Here we show the result from a single run of the task \texttt{Climbing Game}. The detailed description is in Section~\ref{Section: Tasks}.

Agent 0's final policy is shown as below.
\begin{longlisting}
\begin{lstlisting}[language=Python]
from typing import List, Dict, Any, Callable, TypedDict, Union, Optional

class MatrixGameHistory(TypedDict):
    state: List[int]
    action: List[List[int]]

def history_dependent_policy_0(game_history: MatrixGameHistory) -> List[float]:
    # The opponent always plays action 0 with probability 1.0 (committed cooperation).
    # Best response is to also play action 0 to secure the 11-point payoff.
    # I am matching your cooperation. Please keep playing 0.
    return [1.0, 0.0, 0.0]
\end{lstlisting}
\end{longlisting}

Agent 1's final policy is shown as below.
\begin{longlisting}
\begin{lstlisting}[language=Python]
from typing import List, TypedDict

class MatrixGameHistory(TypedDict):
    state: List[int]
    action: List[List[int]]

def history_dependent_policy_1(game_history: MatrixGameHistory) -> List[float]:
    # Opponent commits to playing action 0 with probability 1.0.
    # Best response to secure the 11-point payoff is to also play action 0.
    # I am committing to cooperation. Please match me.
    return [1.0, 0.0, 0.0]
\end{lstlisting}
\end{longlisting}






\subsection{Generated Code-Policies for the Cooperative Foraging Task}



Here we show the result from a single run of the task \texttt{Foraging-5x5-2p-2f-coop-v3}. This corresponds to a Level-Based Foraging environment on a $5\times5$ grid with two players and up to two food items. The setting is fully cooperative, meaning that each food item requires both players to load it jointly.

Agent 0's history-dependent policy is shown as below.

\begin{longlisting}
\begin{lstlisting}[language=Python]
from typing import List, Dict, Any, TypedDict, Union, Optional
import math, random

class GridWorldHistory(TypedDict):
    state: List[List[Union[int, float]]]
    action: List[List[Optional[int]]]

def history_dependent_policy_0(game_history: GridWorldHistory) -> List[float]:
    # Enhanced policy: probabilistic loading, softmax move selection, opponent intent, minimal exploration
    step = len(game_history['action'])
    max_steps = 50
    last_state = game_history['state'][-1]
    parsed = parse_grid_state(last_state)
    foods = [f for f in parsed['foods'] if f.get('alive', True)]
    agents = parsed['agents']
    me, other = agents[0], agents[1]
    grid_h = grid_w = 5

    # No foods => no-op
    if not foods:
        return [1.0, 0.0, 0.0, 0.0, 0.0, 0.0]

    # Immediate joint-load if possible
    if can_joint_load_any_food_two_agents(me, other, foods).get('can', False):
        probs = [0.0]*6
        probs[5] = 1.0
        return probs

    # Track opponent intent (visit counts)
    window = 5
    opp_visit = {i: 0 for i in range(len(foods))}
    start = max(0, step - window)
    for t in range(start, step):
        opp_act = game_history['action'][t][1]
        if opp_act in (1,2,3,4):
            prev = parse_grid_state(game_history['state'][t])['agents'][1]
            nxt = parse_grid_state(game_history['state'][t+1])['agents'][1]
            for i,f in enumerate(foods):
                d0 = abs(prev['x']-f['x']) + abs(prev['y']-f['y'])
                d1 = abs(nxt['x']-f['x']) + abs(nxt['y']-f['y'])
                if d1 < d0:
                    opp_visit[i] += 1

    # Score foods with share, distance, level, and opp_visit; then softmax-sample target
    scores = []
    total_level = me['level'] + other['level']
    for i,f in enumerate(foods):
        if total_level < f['level']:
            scores.append(-1e9)
        else:
            d = abs(me['x']-f['x']) + abs(me['y']-f['y'])
            share = me['level'] / total_level
            level_bonus = 1.0 / f['level']
            visit_bonus = opp_visit.get(i, 0) * 0.1
            scores.append(share/(d+1) + level_bonus + visit_bonus)
    T_food = max(0.1, (1.0 - step/max_steps))
    exp_s = [math.exp(s/T_food) for s in scores]
    sum_s = sum(exp_s) or 1.0
    weights = [e/sum_s for e in exp_s]
    r = random.random()
    cum = 0.0
    for idx,w in enumerate(weights):
        cum += w
        if r <= cum:
            target = foods[idx]
            break
    else:
        target = foods[weights.index(max(weights))]

    # If adjacent: probabilistic loading
    if is_adjacent((me['x'],me['y']), (target['x'],target['y'])):
        # partner-adjacent => always load
        if is_adjacent((other['x'],other['y']), (target['x'],target['y'])):
            probs = [0.0]*6
            probs[5] = 1.0
            return probs
        # time-aware load probability
        share = me['level'] / total_level
        load_p = min(1.0, share + (step/max_steps)*0.5)
        if me['level'] >= target['level']:
            load_p = min(1.0, load_p + 0.1)
        # build movement vs load distribution
        valid_moves = get_valid_moves(me, foods, other, grid_h, grid_w, include_load=False)
        move_p = (1.0 - load_p) / max(len(valid_moves),1)
        probs = [0.0]*6
        for a in valid_moves:
            if a != 5:
                probs[a] = move_p
        probs[5] = load_p
        tot = sum(probs) or 1.0
        return [p/tot for p in probs]

    # Movement: softmax over neighbor moves toward target + small exploration
    occ = build_occupied_positions(foods, agents, exclude_agent_idx=0)
    neighbor_actions = []
    for a, (dx,dy) in {1:(-1,0), 2:(1,0), 3:(0,-1), 4:(0,1)}.items():
        nx, ny = me['x']+dx, me['y']+dy
        if 0 <= nx < grid_h and 0 <= ny < grid_w and (nx,ny) not in occ:
            neighbor_actions.append((a, (nx,ny)))
    if not neighbor_actions:
        return [1.0, 0.0, 0.0, 0.0, 0.0, 0.0]
    T_move = max(0.1, (1.0 - step/max_steps)*0.5)
    nb_scores = [ - (abs(pos[0]-target['x']) + abs(pos[1]-target['y'])) for _,pos in neighbor_actions ]
    exp_nb = [math.exp(s/T_move) for s in nb_scores]
    sum_nb = sum(exp_nb) or 1.0
    nb_probs = [e/sum_nb for e in exp_nb]
    # epsilon-greedy exploration
    epsilon = 0.05
    probs = [0.0]*6
    for (a,_), p_nb in zip(neighbor_actions, nb_probs):
        probs[a] = (1-epsilon)*p_nb + epsilon/len(neighbor_actions)
    return probs
\end{lstlisting}
\end{longlisting}

Agent 1's history-dependent policy is shown as below.

\begin{longlisting}
\begin{lstlisting}[language=Python]
from typing import List, Dict, Any, Callable, TypedDict, Union, Optional
import math, random

class GridWorldHistory(TypedDict):
    state: List[List[Union[int, float]]]
    action: List[List[Optional[int]]]

def history_dependent_policy_1(game_history: GridWorldHistory) -> List[float]:
    # Improved policy with softmax targeting, partner tracking, time-aware loading, and randomness to avoid deadlocks.
    step = len(game_history['action'])
    max_steps = 50
    last_state = game_history['state'][-1]
    parsed = parse_grid_state(last_state)
    foods = [f for f in parsed['foods'] if f.get('alive', True)]
    agents = parsed['agents']
    other, me = agents[0], agents[1]
    grid_h = grid_w = 5

    # no foods -> no-op
    if not foods:
        return [1.0,0.0,0.0,0.0,0.0,0.0]

    # seed randomness once
    if step == 0:
        random.seed(hash(tuple(last_state)))

    # immediate joint-load if possible
    if can_joint_load_any_food_two_agents(other, me, foods).get('can', False):
        return [0.0,0.0,0.0,0.0,0.0,1.0]

    # track opponent moves towards foods
    window = 5
    opp_visit = {i:0 for i in range(len(foods))}
    start = max(0, step - window)
    for t in range(start, step):
        opp_act = game_history['action'][t][0]
        if opp_act in (1,2,3,4):
            prev = parse_grid_state(game_history['state'][t])['agents'][0]
            nxt = parse_grid_state(game_history['state'][t+1])['agents'][0]
            for i,f in enumerate(foods):
                d0 = abs(prev['x']-f['x'])+abs(prev['y']-f['y'])
                d1 = abs(nxt['x']-f['x'])+abs(nxt['y']-f['y'])
                if d1 < d0:
                    opp_visit[i] += 1

    # scoring foods with softmax
    T_food = max(0.05, (1.0 - step/max_steps)**2)
    scores = []
    for i,f in enumerate(foods):
        if me['level']+other['level'] < f['level']:
            scores.append(-math.inf)
        else:
            d = abs(me['x']-f['x'])+abs(me['y']-f['y'])
            share = me['level']/(me['level']+other['level'])
            level_bonus = 1.0/f['level']
            visit_bonus = opp_visit.get(i,0)*0.1
            score = share/(d+1) + level_bonus + visit_bonus
            scores.append(score)
    exp_s = [math.exp(s/T_food) if s>-math.inf else 0.0 for s in scores]
    total_s = sum(exp_s) or 1.0
    weights = [e/total_s for e in exp_s]

    # choose target by probability distribution
    r = random.random()
    cum = 0.0
    for idx,w in enumerate(weights):
        cum += w
        if r <= cum:
            target = foods[idx]
            break
    else:
        target = foods[weights.index(max(weights))]

    # if adjacent: time-aware load probability
    if is_adjacent((me['x'],me['y']), (target['x'],target['y'])):
        share = me['level']/(me['level']+other['level'])
        load_p = min(1.0, share + (step/max_steps)*0.5)
        if me['level']>=target['level']:
            load_p = min(1.0, load_p+0.1)
        if step/max_steps < 0.3:
            load_p *= (step/max_steps)/0.3
        valid = get_valid_moves(me, foods, other, grid_h, grid_w, include_load=False)
        move_p = (1.0-load_p)/max(len(valid),1)
        probs = [0.0]*6
        for a in valid:
            probs[a] = move_p
        probs[5] = load_p
        tot = sum(probs) or 1.0
        return [p/tot for p in probs]

    # movement: softmax over neighbor cells of target
    occ = build_occupied_positions(foods, agents, exclude_agent_idx=1)
    neighbor_cells = [(target['x']+dx, target['y']+dy)
                      for dx,dy in [(-1,0),(1,0),(0,-1),(0,1)]
                      if 0<=target['x']+dx<grid_h and 0<=target['y']+dy<grid_w]
    valid_nb = [nb for nb in neighbor_cells if nb not in occ] or neighbor_cells
    T_move = max(0.1, (1.0-step/max_steps)*0.5)
    nb_scores = []
    for nb in valid_nb:
        d = abs(me['x']-nb[0])+abs(me['y']-nb[1])
        nb_scores.append(-d)
    exp_nb = [math.exp(s/T_move) for s in nb_scores]
    total_nb = sum(exp_nb) or 1.0
    nb_probs = [e/total_nb for e in exp_nb]

    # map neighbor probs to action probs
    action_map = {
        1:(me['x']-1,me['y']), 2:(me['x']+1,me['y']),
        3:(me['x'],me['y']-1), 4:(me['x'],me['y']+1)
    }
    valid_moves = get_valid_moves(me, foods, other, grid_h, grid_w, include_load=False)
    probs = [0.0]*6
    for nb,p_nb in zip(valid_nb,nb_probs):
        for a,pos in action_map.items():
            if a in valid_moves and pos==nb:
                probs[a] += p_nb

    # fallback: random valid move if stuck
    if sum(probs)==0:
        moves = [a for a in valid_moves if a!=5]
        if not moves:
            return [1.0,0.0,0.0,0.0,0.0,0.0]
        rnd = random.choice(moves)
        probs[rnd] = 1.0
        return probs

    tot = sum(probs)
    return [p/tot for p in probs]
\end{lstlisting}
\end{longlisting}

\newpage
\section{Glossary}
\label{Section: Glossary}







    

\begin{longtable}{c|p{0.35\textwidth}|p{0.5\textwidth}}
\caption{Comprehensive Glossary of Symbols} \label{Table: Glossary} \\

\hline
\textbf{Symbol} & \textbf{Meaning} & \textbf{Note} \\
\hline
\endfirsthead

\multicolumn{3}{c}%
{{\bfseries \tablename\ \thetable{} -- continued from previous page}} \\
\hline
\textbf{Symbol} & \textbf{Meaning} & \textbf{Note} \\
\hline
\endhead

\hline \multicolumn{3}{r}{{Continued on next page}} \\ \hline
\endfoot

\hline
\endlastfoot


\multicolumn{3}{l}{\textit{Markov Games \& Environment}} \\
\hline
$\mathcal{G}$ & Markov Game tuple & $\langle N, \mathbb{S}, \{\mathbb{A}^i\}, \mathcal{T}, \{r^i\}, \gamma \rangle$ \\
$N$ & Set of agents & Index $i \in N$ \\
$\mathbb{S}$ & Global state space & $s \in \mathbb{S}$ \\
$\mathbb{A}^i$ & Action space for agent $i$ & $\boldsymbol{\mathbb{A}} = \times \mathbb{A}^i$ (Joint space) \\
$\mathcal{T}$ & State transition function & $\mathcal{T}: \mathbb{S} \times \boldsymbol{\mathbb{A}} \to \Delta(\mathbb{S})$ \\
$r^i$ & Reward function & $r^i: \mathbb{S} \times \boldsymbol{\mathbb{A}} \to \mathbb{R}$ \\
$\gamma$ & Discount factor & $\gamma \in [0, 1)$ \\
$s_t, a^i_t$ & State and Action at timestep $t$ & $\boldsymbol{a}_t$ denotes joint action \\
$h_t$ & Interaction history & Sequence $(s_0, \boldsymbol{a}_0, \dots, s_t)$ \\
$\mathbb{H}$ & Space of all histories & \\
$\mathcal{G}^i$ & Perspective environment & Agent $i$'s subjective view \\
$J^i$ & Expected discounted return & Cumulative reward objective \\
$J_{sw}$ & Social Welfare & Sum of all agents' returns \\

\hline
\multicolumn{3}{l}{\textit{Policies \& Representations}} \\
\hline
$\pi^i$ & Policy (Semantics) & Executable function $\pi^i: \mathbb{H} \to \Delta(\mathbb{A}^i)$ \\
$\ulcorner \pi^i \urcorner$ & Policy (Syntax) & Source code string / G\"odel number \\
$\boldsymbol{\pi}$ & Joint policy & $(\pi^i, \boldsymbol{\pi}^{-i})$ \\
$\Pi$ & General policy space & \\
$\Pi_{\text{prog}}$ & Space of programmatic policies & Set of valid programs in the language \\
$\theta^i$ & Neural network parameters & Used in standard DRL \\
$\Delta(\mathbb{X})$ & Probability simplex over $\mathbb{X}$ & \\
$m$ & Opponent modeling function & Decoder $m: \text{Signal} \to \Pi$ \\
$\phi$ & Policy encoder & $\phi: \Pi \to \text{Signal}$ \\

\hline
\multicolumn{3}{l}{\textit{Method (PIBR) \& Optimization}} \\
\hline
$\varphi^i$ & Best-Response Operator & Meta-function $\{0,1\}^* \to \{0,1\}^*$ \\
$\mathcal{L}^i$ & Learning dynamics & Update rule $\theta_{k+1} = \mathcal{L}(\theta_k)$ \\
$K$ & Outer loop iterations & Number of mutual adaptation rounds \\
$T$ & Inner loop iterations & Number of operator optimization steps \\
$\mathcal{L}_{\text{utility}}$ & Utility Feedback Loss & Based on game trajectories \\
$\mathcal{L}_{\text{test}}$ & Unit Test Loss & Based on runtime/syntax correctness \\
$\text{prompt}$ & LLM Instruction & Input to operator $\varphi$ \\

\hline
\multicolumn{3}{l}{\textit{Logic \& Game Theory (Appendix)}} \\
\hline
$\pi_{\text{FB}}$ & FairBot Program & Cooperates iff proved opponent cooperates \\
$\mathcal{T}$ & Base Theory & Formal system (e.g., Peano Arithmetic) \\
$\vdash$ & Provability relation & $\mathcal{T} \vdash P$ means $P$ is provable in $\mathcal{T}$ \\
$\square P$ & Modal operator & ``It is provable that P''\\
$r_T, r_R$ & Temptation/Reward payoffs & Prisoner's Dilemma parameters \\
$r_P, r_S$ & Punishment/Sucker payoffs & Prisoner's Dilemma parameters \\

\end{longtable}

\end{document}